\documentclass[superscriptaddress,longbibliography,floatfix,twocolumn,nobibnotes]{revtex4-2}
\usepackage{upgreek}
\usepackage{amsmath,braket}
\usepackage{graphicx}
\usepackage{mathtools}
\usepackage{color}
\usepackage{xcolor}
\usepackage{setspace}
\usepackage[pagewise]{lineno}
\usepackage{physics}
\newcommand{\RNum}[1]{\uppercase\expandafter{\romannumeral #1\relax}}


\begin{document}

\title{A chiral one-dimensional atom using a quantum dot in an open microcavity}

\author{Nadia O. Antoniadis}
\email{nadia.antoniadis@unibas.ch}
\affiliation{Department of Physics, University of Basel, Klingelbergstrasse 82, CH-4056 Basel, Switzerland}

\author{Natasha Tomm}
\affiliation{Department of Physics, University of Basel, Klingelbergstrasse 82, CH-4056 Basel, Switzerland}

\author{Tomasz Jakubczyk}
\thanks{Present adress: Faculty of Physics, University of Warsaw, 02-093 Warsaw, Poland}
\affiliation{Department of Physics, University of Basel, Klingelbergstrasse 82, CH-4056 Basel, Switzerland}

\author{R\"{u}diger Schott}
\affiliation{Lehrstuhl f\"{u}r Angewandte Festk\"{o}rperphysik, Ruhr-Universit\"{a}t Bochum, D-44780 Bochum, Germany}

\author{Sascha R. Valentin}
\affiliation{Lehrstuhl f\"{u}r Angewandte Festk\"{o}rperphysik, Ruhr-Universit\"{a}t Bochum, D-44780 Bochum, Germany}

\author{Andreas D. Wieck}
\affiliation{Lehrstuhl f\"{u}r Angewandte Festk\"{o}rperphysik, Ruhr-Universit\"{a}t Bochum, D-44780 Bochum, Germany}

\author{Arne Ludwig}
\affiliation{Lehrstuhl f\"{u}r Angewandte Festk\"{o}rperphysik, Ruhr-Universit\"{a}t Bochum, D-44780 Bochum, Germany}

\author{Richard J. Warburton}
\affiliation{Department of Physics, University of Basel, Klingelbergstrasse 82, CH-4056 Basel, Switzerland}

\author{Alisa Javadi}
\email{alisa.javadi@unibas.ch}
\affiliation{Department of Physics, University of Basel, Klingelbergstrasse 82, CH-4056 Basel, Switzerland}

\date{\today}

\begin{abstract}
In nanostructures, the light-matter interaction can be engineered to be chiral. In the fully quantum regime, a chiral one-dimensional atom, a photon propagating in one direction interacts with the atom; a photon propagating in the other direction does not. Chiral quantum optics has applications in creating nanoscopic single-photon routers, circulators, phase-shifters and two-photon gates. Furthermore, the directional photon-exchange between many emitters in a chiral system may enable the creation of highly exotic quantum states. Here, we present a new way of implementing chiral quantum optics -- we use a low-noise quantum dot in an open microcavity. Specifically, we demonstrate the non-reciprocal absorption of single photons, a single-photon diode. The non-reciprocity, the ratio of the transmission in the forward-direction to the transmission in the reverse direction, is as high as 10.7 dB, and is optimised \textit{in situ} by tuning the photon-emitter coupling to the optimal operating condition ($\beta=0.5$). Proof that the non-reciprocity arises from a single quantum emitter lies in the nonlinearity with increasing input laser power, and in the photon statistics -- ultralow-power laser light propagating in the diode's reverse direction results in a highly bunched output ($g^{(2)}(0)=101$), showing that the single-photon component is largely removed. The results pave the way to a single-photon phase shifter, and, by exploiting a quantum dot spin, to two-photon gates and quantum non-demolition single-photon detectors.
\end{abstract}

\maketitle

In a non-chiral one-dimensional atom, an atom is coupled equally to a right-propagating and to a left-propagating mode in a waveguide. There are two input/output ports, one on the left (port 1) and one on the right (port 2). In the ideal limit (perfect atom with $\beta=1$, where $\beta$ is the probability that the atom emits a photon into the waveguide, a single photon at the input in resonance with the atom), the atom acts as a perfect mirror: the reflectivity is $R=1$; the transmission $T=0$ \cite{auffeves-garnier_giant_2007,shen_theory_2009}. This changes completely in a chiral one-dimensional atom: $R$ and $T$ depend on the propagation direction, left-to-right ($1 \rightarrow 2$) or right-to-left ($2 \rightarrow 1$), i.e.\ the system exhibits non-reciprocity ($T_{1 \rightarrow 2} \ne T_{2 \rightarrow 1}$). There are two simple cases \cite{lodahl_chiral_2017}. First, for $\beta=1$, the atom now becomes perfectly transparent ($T_{1 \rightarrow 2}=1$, $T_{2 \rightarrow 1}=1$, $R_{1 \rightarrow 1}=0$, $R_{2 \rightarrow 2}=0$). In one direction, the $2 \rightarrow 1$-direction, say, the photon is phase-shifted by $\pi$ via the interaction with the atom; in the other direction, $1 \rightarrow 2$, the photon phase-shift is zero. Second, for $\beta=\frac{1}{2}$: in the $2 \rightarrow 1$-direction, the photon is scattered by the atom into non-waveguide modes -- the photon is absorbed -- such that $T_{2 \rightarrow 1}=0$ and $R_{2 \rightarrow 2}=0$, whereas in the $1 \rightarrow 2$-direction, the photon does not interact with the atom, $T_{1 \rightarrow 2}=1$ and $R_{1 \rightarrow 1}=0$. 

Chiral quantum optics has been implemented by using a single emitter in a nano-engineered waveguide, for instance a Rb atom in the evanescent field of a dielectric nanofibre \cite{sayrin_nanophotonic_2015,scheucher_quantum_2016}, or a semiconductor quantum dot in a waveguide \cite{sollner_deterministic_2015,coles_chirality_2016,hurst_nonreciprocal_2018}. In the semiconductor case, $\beta$-factors can be high in nano-beam structures and particularly high in photonic-crystal waveguides. The system becomes chiral provided the quantum dot is located off-centre in a nano-beam \cite{hurst_nonreciprocal_2018}; and at the centre of an inversion-asymmetric photonic-crystal waveguide \cite{sollner_deterministic_2015}. 

We report here a different approach to engineering a chiral one-dimensional atom. A single-mode optical fibre constitutes the waveguide on the ``left" of the quantum dot; another single-mode optical fibre constitutes the waveguide on the ``right" of the quantum dot; the atom itself is a quantum dot in a low-volume one-sided microcavity, where the microcavity is coupled with high efficiency to the single-mode fibres. Chirality is induced by applying a magnetic field to a neutral quantum dot: the quantum dot's $\sigma^{+}$- transition couples to the microcavity and is addressed with $\sigma^{+}$-polarised photons. The advantage of this approach is that the resonant microcavity boosts the light-matter interaction in a controllable way: the $\beta$-factor can be tuned from small to extremely high values (99.7\% has been achieved \cite{najer_gated_2019}). Also, the good mode-matching~\cite{tomm_bright_2021} implies that a high-efficiency, fibre-coupled platform for chiral quantum optics can be constructed. 

\begin{figure*}[t!]
\centering
\includegraphics{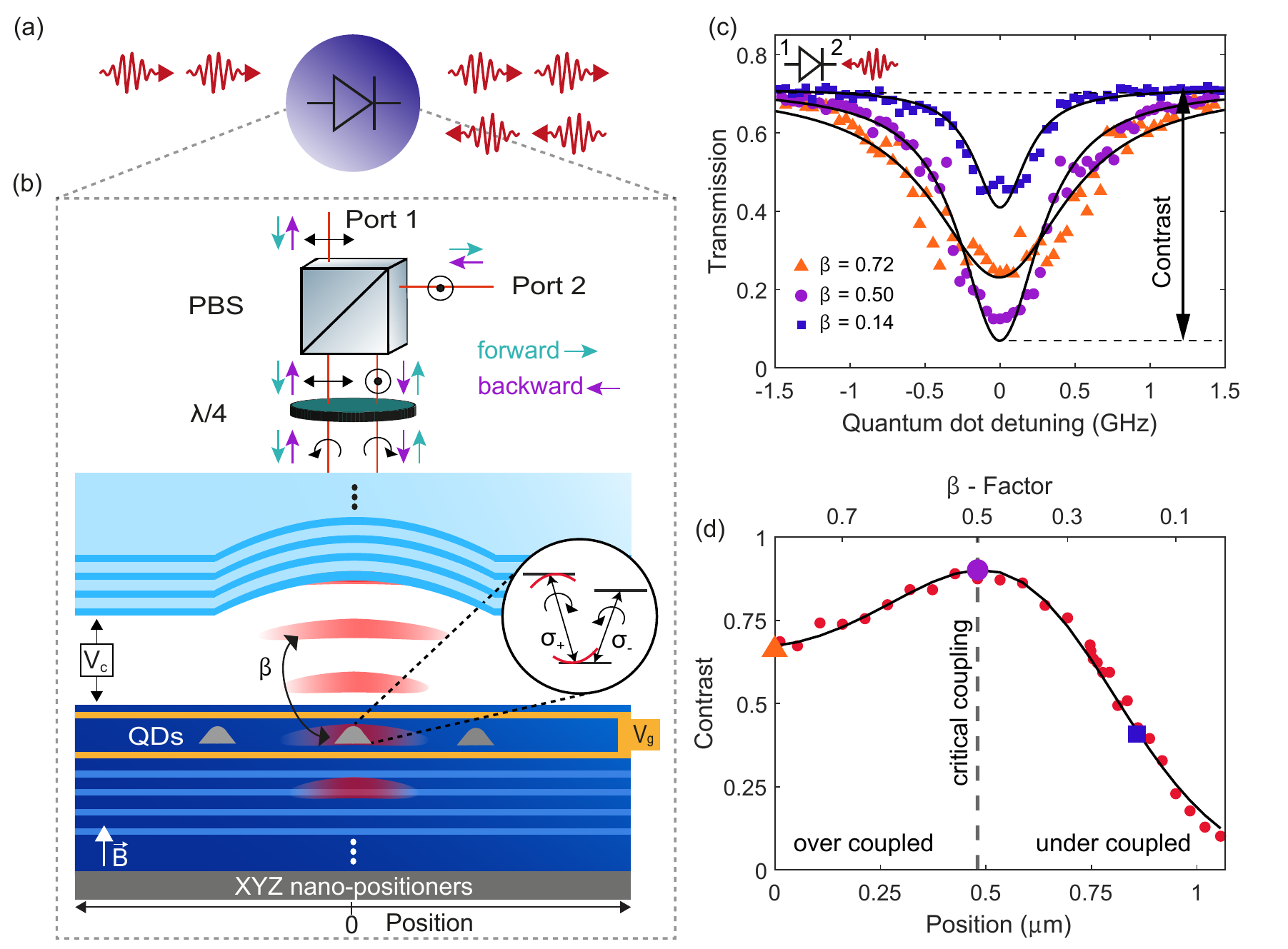}
\caption{\textbf{Schematic and operating principle of the chiral one-dimensional atom}
\textbf{(a)}~The optical system showing the two input/output ports.
\textbf{(b)}~Illustration of the open microcavity. The heterostructure consists of a GaAs/AlAs distributed Bragg reflector (the ``bottom" mirror) and self-assembled InAs quantum dots embedded in an n-i-p diode. Nano-positioners allow precise tuning of both microcavity frequency (via Z) and also the $\beta$-factor by positioning the quantum dots with respect to the anti-node of the microcavity (via XY). An external magnetic field of 2.0~T splits the neutral quantum dot into two circularly-polarised transitions. The polarisation of the light is controlled by a polarising beam-splitter (PBS) and a quarter-wave plate ($\lambda/4$) in the microscope. The $\sigma^{+}$-polarised exciton creates a photon in the microcavity with probability $\beta$. \textbf{(c)}~Transmission from port 1 to port 2 versus quantum dot detuning for three different lateral positions of the quantum dot: orange triangles ($\beta \approx 0.72$), purple circles ($\beta \approx 0.50$) and blue rectangles ($\beta \approx 0.12$). The solid lines are theory curves. The off-resonant transmission is limited to 0.7 by the residual non-degeneracy of the microcavity. The transmission is defined without taking into account the losses in the optical setup. The overall end-to-end throughput of the setup, taking into account all the optical elements and the fibre-coupling, is 56\%: this corresponding to $T = 1$.
\textbf{(d)}~Contrast in the transmission as a function of the lateral position. Coloured data-points indicate the positions used in (b). The black solid line is the theory. 
}
\label{fig1}
\end{figure*}

Here, we implement the chiral scheme with $\beta=\frac{1}{2}$. In one direction, a single photon is transmitted; in the other direction, the photon is absorbed, Fig.~\ref{fig1}a. We call this device a ``single-photon diode" in analogy to its electronic counterpart. The challenge is twofold: to achieve exactly the right $\beta$; and to achieve a close-to-perfect (transform-limited) quantum dot. These challenges were met: we achieve an isolation of 10.7 dB, the highest non-reciprocal response recorded with a single quantum emitter. In addition, the high overall efficiency~\cite{tomm_bright_2021} enables us to observe optical nonlinearities already at an input power of just 100~pW. The quantum nature of this nonlinearity is validated by observation of photon bunching by a factor of 101 compared to that of a laser field. 

\begin{figure*}[t!]
\centering
\includegraphics{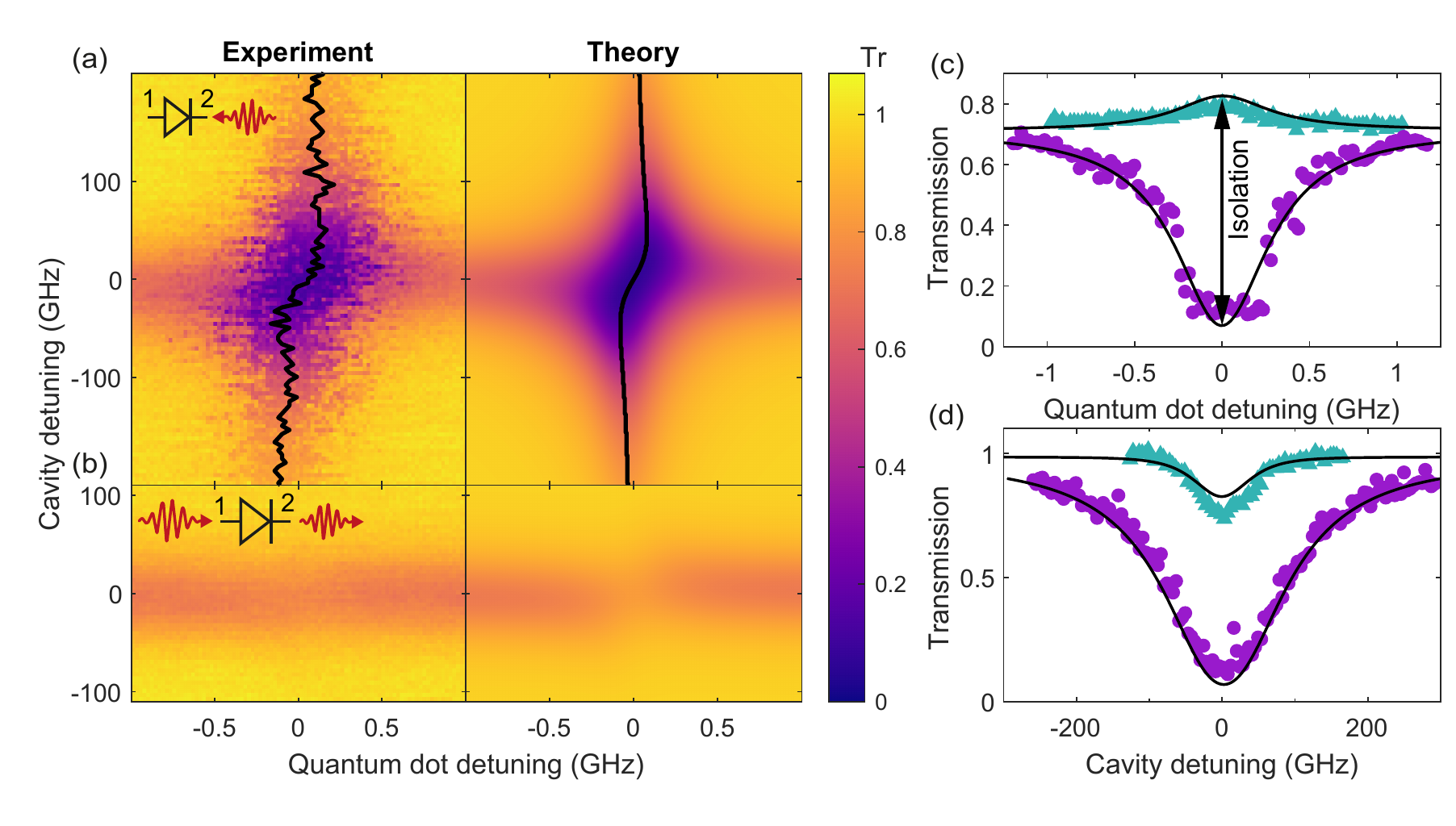}
\caption{\textbf{Non-reciprocal response of the diode.}
Transmission through the system versus quantum dot and microcavity detuning for the backward \textbf{(a)} (port 1 to 2) and forward \textbf{(b)} (port 2 to 1) directions. Experimental data (left) and theory (right) show an excellent match. The black lines show the Lamb shift of the quantum dot resonance induced by the vacuum field of the microcavity.
\textbf{(c)/(d)}~Transmission through the diode on resonance with the microcavity/quantum dot in the forward and the backward direction versus quantum dot/cavity detuning. The purple and turquoise data are cut-throughs of (a) (2$\rightarrow$1) and (b) (1$\rightarrow$2). The black solid lines are the theory
}
\label{fig2}
\end{figure*} 

Figure~1b shows a schematic of the setup and the operation principle of the diode. The optical setup consists of a polarising beam-splitter and a quarter-wave plate set at 45$^{\circ}$~with respect to the polarising beam-splitter axes. Consequently, light propagating in the forward direction is mapped to a left-handed field at the input of the microcavity, while light propagating in the backward direction is mapped to a right-handed field at the microcavity input, thereby creating the spin-momentum locking. The microcavity (Fig.~1b) comprises a highly reflective ``bottom" mirror and a much less reflective ``top" mirror~\cite{tomm_bright_2021}. The top mirror is a dielectric distributed-Bragg-reflector deposited on a crater in a silica substrate with a radius-of-curvature of 11~$\upmu$m. The bottom mirror is a semiconductor distributed-Bragg-reflector on top of which InAs quantum dots are embedded in an n-i-p diode in the heterostructure. The open nature of the microcavity allows the lateral position of the quantum dots to be controlled precisely with respect to the anti-node of the microcavity. This allows \textit{in situ} tuning of the $\beta$-factor. Furthermore, adjusting the distance between the bottom mirror and the top mirror provides precise control over the microcavity's frequency, enabling the quantum dot's $\sigma^{+}$-polarised exciton and the microcavity to be brought into spectral resonance.

We define the transmission as the propagation through the entire diode, i.e.\ from port 1 to 2 (2 to 1) in the forward (backward) direction. Ideally, the transmission in the forward direction $T_{1\rightarrow2}$ is unity as the left-handed optical field is orthogonal to the dipole-moment of the quantum dot, i.e.\ there is no interaction. In the backward direction, the transmission amplitude is given by $t_{2\rightarrow1}=1-2\beta$ where $\beta$ is the $\beta$-factor describing the interaction of the $\sigma^{+}$-polarised exciton with an empty microcavity. The full transmission in the backward direction is $T_{2\rightarrow1}=\left| t_{2\rightarrow1} \right|^2 = \left|1-2\beta \right|^2$. In the over-coupled regime, $\beta\approx1$, the backwards-propagating photons receive a $\pi$-phase shift as they transit through the coupled system. At the critical coupling, $\beta =0.5$, the light reflected by the quantum dot should interfere destructively with the light directly reflected from the microcavity and the transmission through the system vanishes; instead, the photons are scattered into non-microcavity modes.

For this scheme to operate, the microcavity itself should not rotate the polarisation. However, the fundamental microcavity mode is typically split into two modes with orthogonal linear polarisations, a consequence of a birefringence in the semiconductor heterostructure \cite{tomm_tuning_2021}. This mode-splitting decreases with increasing wavelength while the microcavity linewidth, $\kappa/(2\pi)$, increases (Supplementary Information Sec.~\RNum{2}). We operate at a wavelength of 945~nm where the mode-splitting is on the order of 29~GHz and $\kappa/(2\pi)=102$~GHz such that the mode-splitting is much smaller than the linewidth of the microcavity, and the microcavity mode is nearly degenerate. The microcavity losses are dominated by the transmission through the top mirror, i.e.\,$\kappa_{\rm top} \gg \kappa_{\rm bottom}$ (Supplementary Information Sec.~\RNum{2}). Note that in this regime, the residual mode-splitting acts as an overall loss channel and does not limit the non-reciprocity of the system; it mostly affects the insertion loss of the system (here 1.5~dB).

The inset in Fig.~1b shows the level structure of a neutral quantum dot under a magnetic field along the growth direction (Faraday geometry). An out-of-plane magnetic field of 2.0~T splits the right-handed and left-handed transitions by 63~GHz, enough that only one transition interacts with the laser; here, the right-handed dipole $\sigma^{+}$ (see Supplementary Information Sec.~\RNum{2} for characterisation of the quantum dot). The microcavity system resides in a cryostat at a temperature of 4.2~K.

\begin{figure*}[t!]
\centering
\includegraphics{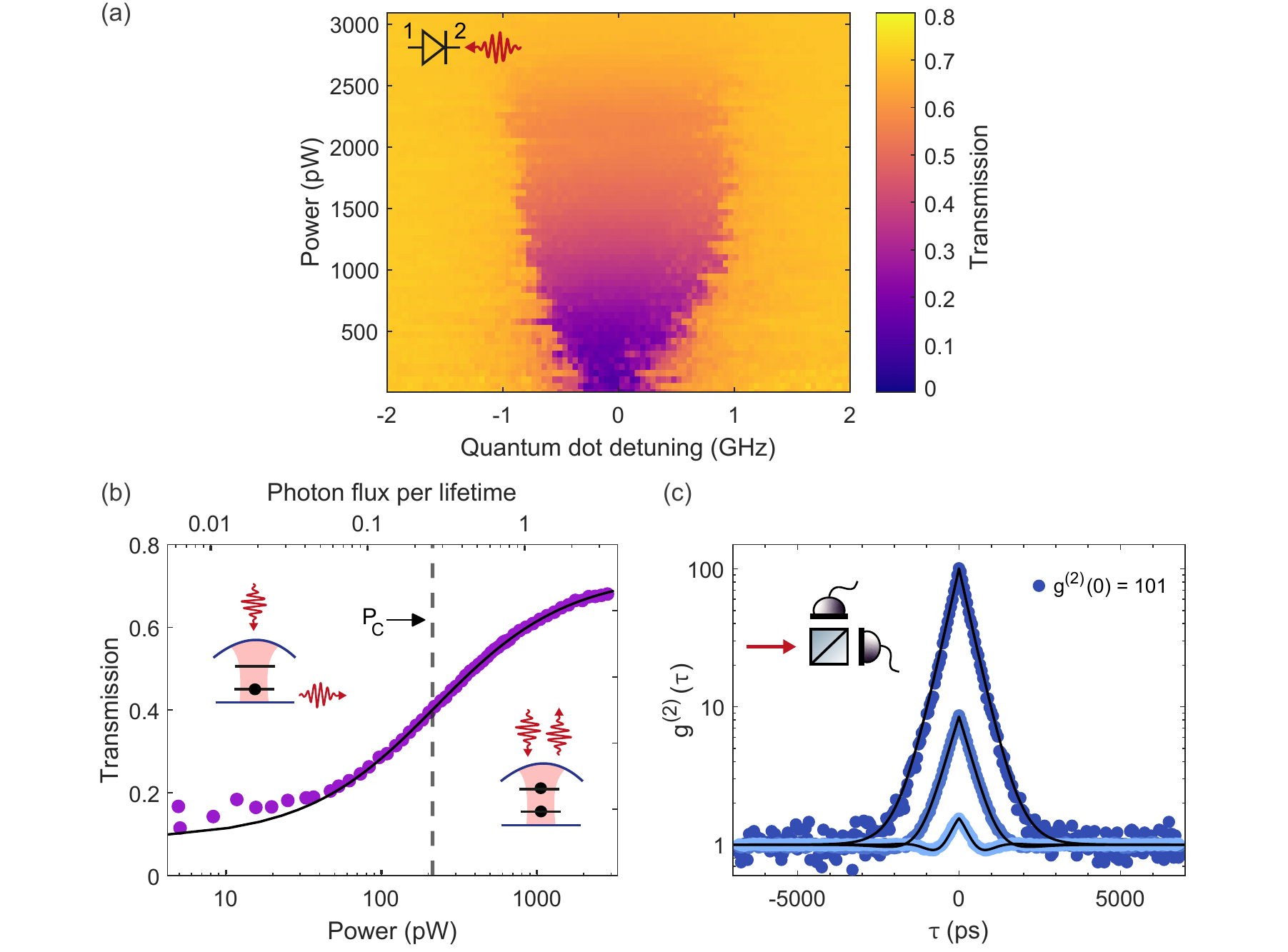}
\caption{\textbf{Nonlinear response of the diode.}
\textbf{(a)} Transmission on resonance with the microcavity versus quantum dot detuning and optical input power. The transmission saturates at 0.7 on increasing the power.
\textbf{(b)} Power dependence of the transmission on resonance with the quantum dot. The black solid line corresponds to the theory. The insets illustrate absorption by the quantum dot at small powers, and transmission at higher powers due to the increasing transparency of the emitter. The critical power is indicated with a vertical line.
\textbf{(c)} Autocorrelation function of the backwards propagating light for three different input optical powers. At the lowest optical power (5~pW) a bunching of 101 is observed. As the power increases the bunching decreases.
}
\label{fig3}
\end{figure*} 

The transmission of the diode is probed with coherent laser light at very low powers, a regime dominated by single-photon components. We exploit the $\beta$-factor's dependence on the lateral position of the quantum dot with respect to the centre of the microcavity to tune the system to the critical-coupling condition. Figure~1c shows the transmission in the backward direction, $T_{2\rightarrow1}$, as a function of the quantum dot frequency for three different positions, i.e.\ three different $\beta$-factors. The orange data points show the transmission for the quantum dot centred in the microcavity. For this position we extract $\beta=0.72$, matching well the theoretical expectations for the maximum achievable $\beta$ at this wavelength (Supplementary Information Sec.~\RNum{3}). The $\beta$-factor is evaluated via the linewidth and the contrast of the dip. When the quantum dot is laterally displaced such that $\beta=0.50$ (purple data points), the $T_{2\rightarrow1}$ is lowest: $T_{2\rightarrow1}=0.07$. Also, the linewidth of the quantum dot transition decreases with decreasing $\beta$ due to a reduced Purcell-enhancement. We define the transmission contrast as one minus the ratio between transmission on resonance and the transmission far off-resonance with the quantum dot ($1-\frac{T_{2\rightarrow1}^0}{T_{2\rightarrow1}^\infty}$). The transmission contrast is measured while scanning the lateral position of the quantum dot relative to the microcavity's optical axis (Fig.~1d). For a well-centred position, the contrast in the transmission is around 0.67, increasing to a value of 0.9 as $\beta$ approaches 0.5, and decreasing as $\beta$ is further reduced. The solid black lines in Fig.~1c,d are the theoretically expected behaviour. (A comprehensive model for transmission through a two-level system coupled to a non-degenerate one-sided microcavity is derived and discussed in detail in Supplementary Information Sec.~\RNum{1}.) An average spectral fluctuation of the quantum dot of 40~MHz was found by comparing the theoretical model and the measurements \cite{kuhlmann_charge_2013}. 

For the operation of the single-photon diode, we focus at the position with the largest transmission contrast, $\beta=0.50$. The non-reciprocal nature of the diode is demonstrated by measuring the transmission in both directions at the critical-coupling condition. Figure~2a shows the transmission through the diode in the backward direction as a function of the microcavity detuning and the quantum dot detuning. The transmission contrast shown in these maps decreases with microcavity detuning on account of a reduced $\beta$-factor. On the contrary, the transmission in the forward direction (Fig.~2b) presents an almost flat behaviour independent of the detuning from the quantum dot's resonance. The panels on the right side of Fig.~2a,b show the theoretically predicted behaviour from our model (Supplementary Information Sec.~\RNum{1}). A comparison of the transmission in the forward and backward directions as a function of the quantum dot (microcavity) detuning is shown in Fig.\,2c(d). At resonance, the transmission in the forward direction is around 0.82. The slight increase in the transmission signal in the forward direction is attributed to the mode-splitting. A figure of merit for a diode is the isolation. It is defined as $T_{1\rightarrow2}/T_{2\rightarrow1}$ and found to be a factor 11.9 (corresponding to 10.7~dB). 

To prove that the non-reciprocity arises from a single emitter, we probe both the power dependence and the photon statistics of the output. In the power dependence, we find a striking nonlinearity of the transmission in the backward direction. Figure.~3a shows the transmission in the backward direction as a function of the optical power and detuning from the quantum dot's resonance. On resonance with the quantum dot (Fig.~3b), the backward transmission increases with a power-law dependence with a slope of one and is described by $T_{2\rightarrow1}^0=\frac{P/P_\textrm{C}}{1 + P/P_\textrm{C}}$, where $P$ is the input optical power and $P_\textrm{C}$ the critical power. The experimental data in Fig.~3b match this behaviour very well for $P_\textrm{C} \,=\, 213$~pW. This behaviour is characteristic of the saturation of a two-level system -- while the interaction between the input field and the quantum dot is linear in power at very low powers, the quantum dot saturates at higher input powers, which leads to a strong power-dependent transmission. The critical power is very close to the theoretically expected value of 198~pW (Eq.~9 in Supplementary Information Sec~\RNum{1}). This power level corresponds to an average photon flux of $\braket{n}=0.27$ at the input of the microcavity per lifetime of the quantum dot ($\tau_\textrm{QD}\,=\,0.26$~ns). 

The very low onset of the nonlinearity implies that the quantum statistics of the output field are affected by interaction with the quantum dot \cite{rice_single-atom_1988}. We verify this by measuring the second-order auto-correlation function, $g^{(2)}(\tau)$, of the backward transmitted light \cite{wang_turning_2019-1,pscherer_single-molecule_2021,tiecke_nanophotonic_2014, snijders_purification_2016}. The $g^{(2)}(\tau)$ was measured for three different powers (Fig.~3c). At the lowest power (5~pW), a very strong bunching of 101 is observed, proving that the single-photon components of the laser have been largely removed by the quantum dot. With increasing power, the bunching decreases rapidly and eventually vanishes, corresponding exactly to the expected behaviour on saturating the quantum dot: at high powers, most of the laser light is transmitted without interaction, resulting in $g^{(2)}(\tau)=1$, the auto-correlation function of the laser light. Additionally, the auto-correlation function of the transmission in the forward direction is constant and unitary -- this confirms the non-reciprocal transmission in the system (see Supplementary Information Sec.~\RNum{4}). The measurements were modelled (see Supplementary Information Sec.~\RNum{4}) and the results are depicted as a solid black line in Fig.~3c. We emphasise that all the data in Fig.~2 and Fig.~3 are modelled with the same set of parameters, in particular $\beta=0.50$ and a free-space decay rate $\gamma_0/(2\pi)=300$~MHz.

The reported experiments reveal a strong non-reciprocal and highly nonlinear transport of optical photons through a quantum dot-microcavity system. Model calculations based on the canonical chiral one-dimensional atom describe the system extremely well. The single-photon diode is realised with a modest emitter-microcavity coupling, $\beta=0.5$. We can foresee a range of applications. For example, the non-reciprocal behaviour can be dynamically controlled by driving the $\sigma^{-}$ transition, by using the spin-state of a charge carrier in the quantum dot, or by fast Stark tuning of the quantum dot, opening possibilities for optical switches and transistors \cite{chang_single-photon_2007, witthaut_photon_2012, javadi_spinphoton_2018,shomroni_all-optical_2014, oshea_fiber-optical_2013}. Theory predicts that the strong bunching of the photons in the transmission of the system presages the formation of a two-photon bound state -- it is a first step in creating exotic photonic states and simulating many-body dynamics using photons \cite{noh_quantum_2016, hafezi_non-equilibrium_2013, gullans_efimov_2017, maghrebi_coulomb_2015}. The performance of the system can be further improved by eliminating the mode-splitting of the microcavity. The mode-splitting can be minimised, perhaps eliminated, by exploiting the electro-optic effect \cite{frey_electro-optic_2018} or by applying uni-axial stress to the semiconductor heterostructure \cite{tomm_tuning_2021}. This would not only reduce the insertion losses but also bring the regime $\beta=1$ within range. Such a device would be ideal for achieving a single-photon phase-shifter and has a strong potential for deterministic two-photon quantum gates, either by using spin-state of the quantum dot or by exploiting photonic bound states \cite{ralph_photon_2015, mahmoodian_dynamics_2020}. 

\section*{Acknowledgements}\vspace{-3mm}
We thank Sahand Mahmoodian and Daniel Najer for stimulating discussions. The work was supported by NCCR QSIT and SNF Projects No.s 200020$\_$175748 and 200020$\_$204069. AJ received funding from the European Unions Horizon 2020 Research and Innovation Programme under the Marie Sk\l odowska-Curie grant agreement No.\ 840453 (HiFig). TJ acknowledges support from the European Unions Horizon 2020 Research and Innovation Programme under the Marie Sk\l odowska-Curie grant agreement No.\ 792853 (Hi-FrED). RS, ADW and AL acknowledge financial support from the grants DFH/UFA CDFA05-06, DFG TRR160, DFG project 383065199, and BMBF Q.Link.X 16KIS0867. \\ \\

\newpage 

\onecolumngrid 
\newpage 
\section*{Supplementary Information: \\ A chiral one-dimensional atom using a quantum dot in an open microcavity}

\section{\vspace{-0.5cm}Transmission of a two-level emitter coupled to a one-sided cavity}

\begin{figure}[t!]
    \centering
    \includegraphics[width=\linewidth]{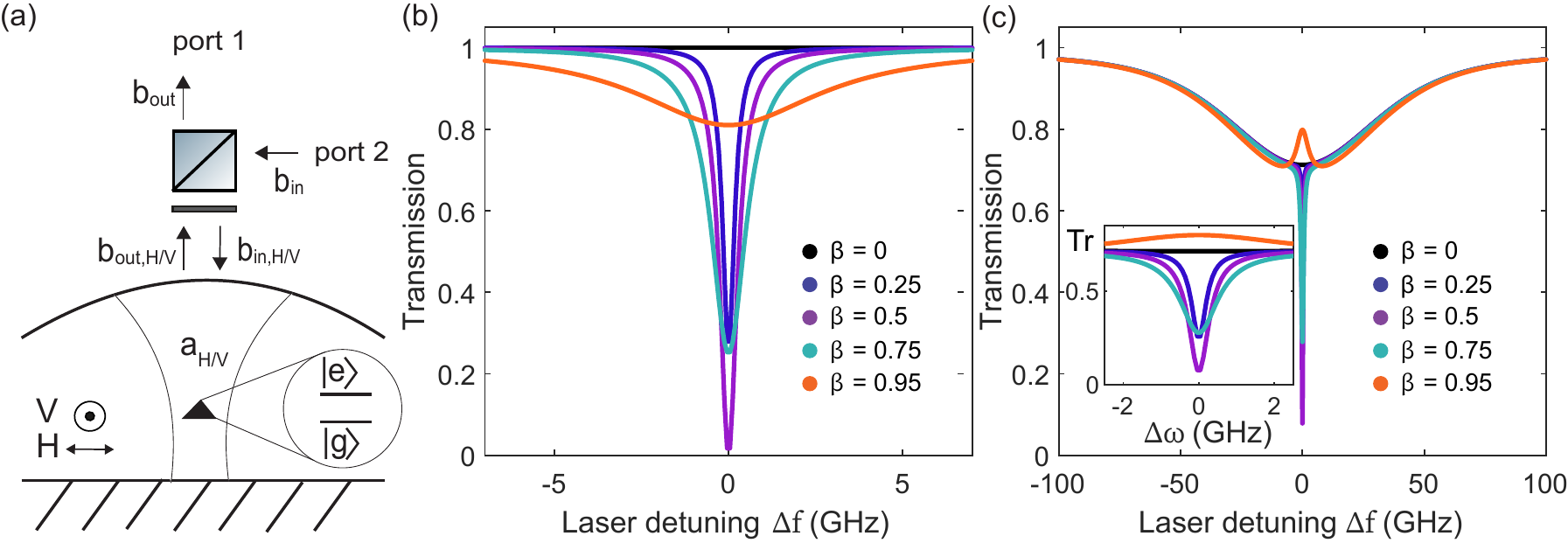}
    \caption{\textbf{(a)} Illustration of the concept and definition of the parameters of a one-sided cavity with cavity modes $a_\textrm{H}$ and $a_\textrm{V}$, and incoming (outgoing) field $b_{\textrm{in}}(b_{\textrm{out}})$. A two-level emitter with ground state $\ket{g}$ and excited state $\ket{e}$ is coupled to the cavity modes.
    \textbf{(b)} Transmission through the full system in (a) as a function of the laser frequency detuning, $\Delta f = \Delta\omega/(2\pi)$, for different $\beta$-factors in the simplified, ideal case: $\delta_\textrm{H}/(2\pi)=\delta_\textrm{V}/(2\pi)=0$, without spectral fluctuation, $\kappa/(2\pi)=$~102\,GHz and $\gamma/(2\pi)=$\,0.30\,GHz. 
    \textbf{(c)} Transmission through the full system in (a) as a function of the laser frequency detuning, $\Delta f = \Delta\omega/(2\pi)$, for different $\beta$-factors under experimental conditions: $\delta_\textrm{H}/(2\pi)-\delta_\textrm{V}/(2\pi)=$\,29\,GHz, $\delta_{\textrm{sf}}/(2\pi)=40$\,MHz, $\kappa/(2\pi)=$~102\,GHz and $\gamma/(2\pi)=$\,0.30\,GHz. The inset shows how the depth of the dip and the linewidth depend on the $\beta$-factor. } 
    \label{fig:scetch}
\end{figure}

The full model describing the transmission of a two-level system coupled to a one-sided cavity is derived based on the theory in Ref.\,\cite{auffeves-garnier_giant_2007}. We consider the situation in Fig.\,\ref{fig:scetch}a. The cavity has two orthogonally polarised modes (H and V). Both modes are coupled to a one-dimensional waveguide. The Hamiltonian for the system is:
\begin{align}
\begin{split}
    \textrm{H} = ~& \hbar\omega_0\sigma_z+\hbar(\omega_0+\delta_\textrm{H})a_\textrm{H}^\dagger   a_\textrm{H}+\hbar(\omega_0+\delta_\textrm{V})a_\textrm{V}^\dagger a_\textrm{V}\\ 
    &+ i\hbar g_\textrm{H}(\sigma_+a_\textrm{H}-\sigma_-a_\textrm{H}^\dagger)+i\hbar(g_\textrm{V}^*\sigma_+a_\textrm{V}-g_\textrm{V}\sigma_-a_\textrm{V}^\dagger) \\ 
    &+ \sum_{k}\hbar\omega b_\textrm{kH}^\dagger b_\textrm{kH}+\sum_{k}\hbar\omega b_\textrm{kV}^\dagger b_\textrm{kV}\\
    &+ \sum_{k}\hbar \kappa_- (a_\textrm{V}b_\textrm{kH}^\dagger+a_\textrm{V}^\dagger b_\textrm{kH})+\sum_{k}\hbar \kappa_+ (a_\textrm{V}b_\textrm{kV}^\dagger+a_\textrm{V}^\dagger b_\textrm{kV}) 
    \end{split}
    \label{eq:hamiltonian}
\end{align}
The cavity contains a single two-level system (TLS) with angular resonance frequency $\omega_0$. $a_\textrm{H/V}$ ($b_\textrm{kH/kV}$) describes the annihilation operator for the cavity mode (waveguide fields). $\omega_0 + \delta_\textrm{H/V}$ is the cavity angular frequency, $\omega$ is the laser angular frequency and $\Delta\omega = \omega_0-\omega$. The atomic operators are given by $\sigma_z = \frac{1}{2}(\ketbra{e}{e}-\ketbra{g}{g})$ and $\sigma_- = \ketbra{g}{e}$. The coupling strengths between the cavity modes and the TLS are given by $g_\textrm{H,V}$ where $g_\textrm{V}$ can be complex. 
Applying the same procedure as Ref.\,\cite{auffeves-garnier_giant_2007}, we end up with the following equations of motion for the cavity modes and the atomic operators in the laser frame: 
\begin{align}
\begin{split}
    \dot{a}_{\textrm{H/V}} &= -i\left(\Delta \omega +\delta_{\textrm{H/V}}\right) \cdot a_{\textrm{H/V}} - \frac{\kappa}{2} a_\textrm{H/V} - g_\textrm{H/V}\cdot \sigma_- + i\sqrt{\kappa} \cdot b_{\textrm{in,H/V}}, \\
    \dot{\sigma}_- &= -i\Delta\omega \sigma_- -2g_\textrm{H}\cdot \sigma_z \cdot a_H -2g^*_\textrm{V}\cdot \sigma_z \cdot a_V -\gamma/2\cdot \sigma_-, \\
    \dot{\sigma}_z &= g_\textrm{H} \cdot \left(\sigma_+ a_H + a_H^\dagger \sigma_-\right) + \left(g^*_\textrm{V}\sigma_+ a_V + g_\textrm{V}a_V^\dagger \sigma_-\right) -\gamma \cdot \left(\sigma_z +\frac{1}{2}\right), \\
    b_{\textrm{out,H/V}} &= b_{\textrm{in,H/V}} + i \sqrt{\kappa} \cdot a_{\textrm{H,V}},
        \end{split}
    \label{eq:original}
\end{align}
where $\gamma$ is the decay rate of the two-level system into non-cavity ``leaky" modes and is introduced using a Lindblad operator, and $\kappa$ is the photon decay rate through the top mirror. $b_\textrm{in,H/V}$ describe the horizontally/vertically polarised components of the input field, and $b_\textrm{out,H/V}$ describe the corresponding output fields.

Equations \,\ref{eq:original} are the quantum coupled mode equations for the evolution of a TLS and a cavity driven by $b_{\textrm{in}}$. In the ``bad cavity" regime ($\kappa \gg g$), the cavity mode can adiabatically be eliminated from the equations by setting $\dot{a} = 0$. This implies: 
\begin{align}
\begin{split}
    a_{\textrm{H,V}} &= \frac{i\sqrt{\kappa} \cdot b_{\textrm{in,H,V}} - g_{\textrm{H,V}} \cdot \sigma_-}{\frac{\kappa}{2} + i (\Delta \omega + \delta_{\textrm{H,V}})}.\\
    \label{eq:cavity}
\end{split}
\end{align}
The linearly-polarised field at the input of the diode (the diode consists of the cavity, the TLS and the optical components, as shown in Fig.\,\ref{fig:scetch}), $b_\textrm{in}$, is converted to a circularly polarised field at the input of the cavity, and $b_\textrm{in,H}=1/\sqrt{2}b_\textrm{in}$ and $b_\textrm{in,V}=i/\sqrt{2}b_\textrm{in}$. Moreover, the TLS has a circularly-polarised optical dipole, hence,
$g_\textrm{V}=ilg_\textrm{H}$, where $l=\pm1$ sets the handedness of the optical dipole. In the following, we drop the subscript H, setting $g_\textrm{H}=g$.
Equation \ref{eq:cavity} may be written in a more compact form by using the definition $t_{\textrm{H,V}} = \left(1+{2i \left(\Delta \omega + \delta_{\textrm{H,V}}\right)}/{\kappa} \right)^{-1}$:
\begin{align}
\begin{split}
    a_{\textrm{H}} &= \sqrt{\frac{2}{\kappa}}\left(ib_{\textrm{in}} - \sqrt{\frac{2g^2}{\kappa}} \sigma_-\right)t_\textrm{H},\\
    a_{\textrm{V}} &= i\sqrt{\frac{2}{\kappa}}\left(ib_{\textrm{in}} -l \sqrt{\frac{2g^2}{\kappa}} \sigma_-\right)t_\textrm{V}.\\
    \label{eq:cavitycompact}
\end{split}
\end{align}
The output of the cavity passes a quarter wave-plate and is converted to a vertical field at the output of the diode, and hence, $b_\textrm{out}=\frac{1}{\sqrt{2}} (b_\textrm{out,H}-ib_\textrm{out,V})$. Combining this equation with the last term in Eq.~\ref{eq:original}, the field amplitude at the output of the diode can be related to the input fields and the cavity operators as:
\begin{align}
    \begin{split}
    b_\textrm{out}&=b_\textrm{in}+\sqrt\frac{\kappa}{2}(ia_\textrm{H}+a_\textrm{V}),\\
    &=b_\textrm{in}-b_\textrm{in}(t_\textrm{H}+t_\textrm{V})-i\sqrt\frac{\Gamma_1}{2}(t_\textrm{H}+lt_\textrm{V})\cdot\sigma_-,
        \end{split}
     \label{eq: inoutformula}
\end{align}
and the equations motion become: 
\begin{align}
    \begin{split}
    \dot{\sigma}_-  &= - i \Delta \omega \sigma_- -\frac{\Gamma_1}{2}\left(t_H + t_V + \frac{1}{F_{p1}}\right)~\sigma_- + ib_\textrm{in} \sqrt{\frac{\Gamma_1}{2}} (-2\sigma_z)(t_H +lt_V),  \\
    \dot{\sigma}_z &= -{\Gamma_1}\left(\textrm{Re}(t_H +l t_V)+\frac{1}{F_{p1}}\right)\left(\sigma_z + \frac{1}{2}\right) + b_\textrm{in}\sqrt{\frac{\Gamma_1}{2}} \left(i(t_H + t_V)\sigma_+ + h.c.\right) 
     \label{eq: pluggedcav3},
        \end{split}
\end{align}
where $\Gamma_1 = \frac{4g^2}{\kappa}$, $F_{p1} = \frac{4g^2}{\kappa \gamma}$. Note that the decay rate of the TLS is $\Gamma=2\Gamma_1$ when the cavity modes are degenerate. In this limit, the Purcell-factor is $F_p=2F_{p1}$ as both cavity modes contribute to the total Purcell-factor.

Next, we solve the steady-state equations $\dot{\sigma}_- = \dot{\sigma}_z = 0$ for $\sigma_z$ and $\sigma_-$. Following the Ehrenfest theorem, we can replace the operators in the steady state by their expectation values:\,$\expval{\sigma_-}$ = $S_-$, $\expval{\sigma_z}$\,=\,$S_z$ and $\expval{b_{\textrm{in}}}$\,=\,$b_{\textrm{in}}$. We end up with:
\begin{align}
    \begin{split}
    S_- &= \frac{ib_\textrm{in}\sqrt{\frac{\Gamma}{4}}(-2S_z)(t_\textrm{H}+lt_\textrm{V})}{-i\Delta\omega+\Gamma/4\left(t_H + t_V + \frac{2}{F_{p}}\right)~},\\
    S_z&= \frac{-1/2\left(\textrm{Re}(t_H +l t_V)+\frac{2}{F_{p}}\right)}{\textrm{Re}(t_H +l t_V)+\frac{2}{F_{p}}+2b^2_\textrm{in}\textrm{Re}[\frac{(t_\textrm{H}+t_\textrm{V})(t^*_\textrm{H}+lt^*_\textrm{V})}{i\Delta\omega+\Gamma/4(t^*_\textrm{H}+t^*_\textrm{V}+2/F_p)}]}. 
    \label{eq:S-}
        \end{split}
\end{align}
For the simplified case of resonant excitation ($\Delta\omega = 0$) and degenerate cavity modes ($\delta_H = \delta_V$), $S_z$ reduces to:
\begin{equation}
   S_z = -\frac{1}{2}\frac{1}{1+|b_{\textrm{in}}|^2/P_c} = -\frac{1}{2}\Pi,
\end{equation}
with critical power:
\begin{equation}
    P_c = \frac{\Gamma}{8\beta^2} \cdot \hbar \omega.
    \label{eq: powertheo}
\end{equation}
$\beta$ is the coupling efficiency of the emitter to the cavity and can be described by: $\beta=\frac{\Gamma}{\Gamma+\gamma}=\frac{F_p}{F_p+1}$.
Plugging Eq.\,\ref{eq:S-} into Eq.\,\ref{eq: inoutformula}, we determine the full transmission coefficient of the cavity-emitter system via $b_{\textrm{out}} = t \cdot b_{\textrm{in}}$ to be: 
\begin{equation}
    t =~1~-~t_H~-t_V~+~\frac{\Pi}{\frac{2 \Delta\omega}{\frac{\Gamma}{2} (t_H + t_V)} + 1 + \frac{2}{F_p (t_H + t_V)}}\cdot\frac{(t_H + lt_V)^2}{(t_H+t_V)}. 
    \label{eq:refl_tot}
\end{equation}
When $l=1$, the optical dipole and the incoming light have the same handedness, while $l=-1$ for a non-interacting optical dipole. At low input powers, with the same simplified condition as in Eq.\,8 ($\Delta\omega = 0$, $\delta_H = \delta_V$) and $l=1$, the reflection coefficient can be reduced to the simple form:
\begin{equation}
    t(0)~= -1+2\beta. 
    \label{eq:refl_resonance}
\end{equation}
In this case, $t(0) = 0$ at the critical coupling condition $\beta = 1/2$.

\newpage
So far, $t$ includes only the coherent part of the transmitted field. At higher excitation powers, the transmitted field is a sum of coherent and incoherent components. The incoherent component arises from the spontaneous emission of photons from the TLS. The incoherent part of the scattering ($P_{\textrm{incoh}}$) can be extracted by subtracting the coherent scattering of the TLS (both scattered into the cavity and into non-cavity modes) from the total power \cite{javadi_single-photon_2015}:
\begin{equation}
    \frac{P_{\textrm{incoh}}}{|b_{\textrm{in}}|^2} = 1 - |t|^2 - \gamma \cdot \frac{|\expval{S_-}|^2}{|b_{\textrm{in}}|^2} .
\end{equation}
The intensity transmission from port 1 to port 2 (or vice versa) in Fig. 1b of the main text is given by
\begin{equation}
    T = |t|^2 + \beta \cdot \frac{P_{\textrm{incoh}}}{2 \cdot |b_{\textrm{in}}|^2}. 
    \label{eq:transm}
\end{equation}
In the simplest ideal case (resonance condition as in Eq.\,8 and $\beta = 0.5$), this leads to a  power-dependent transmission of the form:
\begin{equation}
T =  \frac{{|b_{\textrm{in}}|^2}/{P_c}}{(1+{|b_{\textrm{in}}|^2}/{P_c})}.
\label{eq:powersimple}
\end{equation}
The average photon flux per lifetime of the emitter is related to the input power via $\hbar\omega\expval{n} = |b_{\textrm{in}}|^2/(\Gamma+\gamma)$.

In order to match the conditions of the experiment, we include a spectral fluctuation. This is done by convoluting a Lorentzian distribution with the quantum dot transmission \cite{kuhlmann_charge_2013}:
\begin{equation}
   T_{\textrm{end}} = \int T~(\Delta\omega_{\textrm{QD}}+\sigma) \cdot L(\sigma, \delta_{\textrm{sf}}) \cdot d\sigma.
    \label{eq:diff}
\end{equation}
$L(\sigma, \delta_{\textrm{sf}})$ describes a Lorentzian distribution and $\delta_{\textrm{sf}}$ is the FWHM of this distribution, i.e.\ the spectral fluctuation. \\

Figures \ref{fig:scetch}b and c show the transmission through the system as a function of the laser frequency for different values of the $\beta$-factor. We use $\kappa/(2\pi)=102$\,GHz and $\gamma/(2\pi)=0.30$\,GHz for both figures. In Fig.\,\ref{fig:scetch}b we show the simplified, ideal case without mode-splitting and without spectral fluctuation ($\delta_\textrm{H}/(2\pi)=\delta_\textrm{V}/(2\pi)=$0, $\delta_{\textrm{sf}}/(2\pi)=0$). The transmission is unity in the case of zero coupling between TLS and cavity ($\beta=0$). With increasing $\beta$, a dip appears on resonance with the TLS. This dip reaches zero at the critical coupling condition $\beta=0.5$. Increasing $\beta$ further reduces the depth of the dip again. In the strongly over-coupled regime ($\beta\approx1$), the photons that interact with the TLS acquire a $\pi$-phase shift according to Eq.\,\ref{eq:refl_resonance}. Figure\,\ref{fig:scetch}c shows the same plot after adapting the theory to the experimental conditions, $\delta_{\textrm{sf}}/(2\pi)=40$\,MHz and $\delta_\textrm{H}/(2\pi)-\delta_\textrm{V}/(2\pi)=$\,29\,GHz.
With no TLS-cavity coupling ($\beta=0$), the transmission through the system is governed by the two cavity modes and a dip of 30\% arises due to the finite mode-splitting of the cavity. In the under-coupled regime ($\beta< 0.5$), a small dip is visible on resonance with the quantum dot, similar to the ideal case. This dip reaches 90\% for the critical-coupling condition ($\beta=0.5$). At the critical-coupling condition, most of the photons impinging on the system are dissipated by the TLS to the non-cavity modes. Increasing $\beta$ further results in a reduced dip size.  

We model the transmission plots in the main text using Eq.~\ref{eq:diff} with the same parameters as in Fig.\,\ref{fig:scetch}. For the results in Fig.\,2 and Fig.\,3b of the main text, we set $F_p=1$. The saturation power extracted by fitting Fig.\,3b in the main text with Eq.\,\ref{eq:powersimple} is $P_c=213$\,pW, which is in agreement within the errorbar with the theoretical saturation power $P_c=198$\,pW that is expected from Eq.\,\ref{eq: powertheo}.


\newpage

\section{Microcavity and quantum dot characterisation \label{sec:modesplitting}}
\subsection{Characterisation of the cavity parameters}
By measuring the cross-polarised~\cite{kuhlmann_dark-field_2013} transmission through the bare cavity setup (from port 2 to port 1 in Fig. \ref{fig:scetch}a) and fitting a double-Lorentzian to the data, we can characterise the decay rate, $\kappa$, and the mode-splitting, $\Lambda = \delta_H - \delta_V$, of the cavity; see the top panel in Fig.\,\ref{fig:cavity}b for an example of the transmission spectrum. The top panel in Fig.\,\ref{fig:cavity}a shows $\kappa$ and $\Lambda$ as a function of wavelength; $\Lambda$ decreases for higher wavelength while $\kappa$ increases~\cite{tomm_tuning_2021}. By comparing $\kappa$ to the cavity losses reported in Ref.\,\cite{najer_suppression_2021}, which were measured with the same bottom mirror but with a highly reflective top mirror, we conclude that $\kappa\approx\kappa_{\textrm{top}}$ ($\kappa_{\textrm{top}}$ describes the cavity losses through the top mirror), i.e.\ close to all losses arise via transmission through top mirror for all wavelengths, assuring that the cavity is indeed one-sided in the full range of wavelengths presented.

\subsection{Characterisation of the insertion loss of the diode and the optimal wavelength for its operation}
\begin{figure}[b!]
    \centering
    \includegraphics{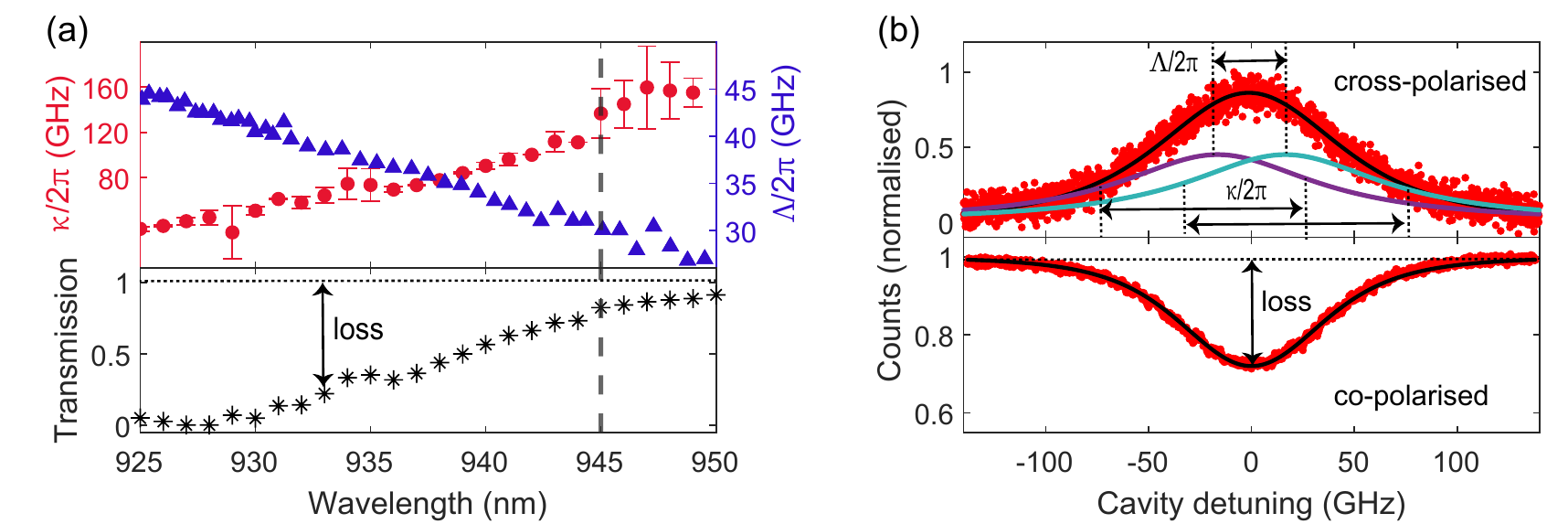}
    \caption{\textbf{Characterisation of the cavity parameters} without a QD (a) Measured cavity decay rate ($\kappa/(2\pi)$, red circles) and mode-splitting ($\Lambda/(2\pi)$, blue triangles) versus wavelength. The transmission through the one-sided cavity on resonance with the cavity mode (black stars) is calculated with these two parameters and Eq.\,\ref{eq:cav_tr}. The transmission increases, i.e.\ the loss decreases, with increasing wavelength. (b) The measured transmission through the cavity mode in the cross- (top) and co-polarised (bottom) configuration at a wavelength of 945\,nm versus the detuning of the cavity frequency. The peak in cross-polarisation is fitted with the response of the two linear cavity modes (H and V) that are plotted in purple and turquoise. The cavity dip was modelled using the parameters from (a).}
    \label{fig:cavity}
\end{figure}

Neglecting interaction terms in Eq.\,\ref{eq:refl_tot}, we can calculate the transmission through both cavity modes in the co-polarised configuration as:
\begin{equation}
    T = \left| 1 - \frac{1}{1 + {2i(\Delta \omega + {\Lambda}/{2})}/{\kappa}} - \frac{1}{1+{2i(\Delta \omega -{\Lambda}/{2})}/{\kappa}} \right|^2.
    \label{eq:cav_tr}
\end{equation}
Using Eq.\,\ref{eq:cav_tr} and $\kappa$ and $\Lambda$ extracted in Sec.\,\ref{sec:modesplitting}A (shown in the top panel of Fig.\,\ref{fig:cavity}a), we calculate the transmission from port 2 to port 1 for circularly polarised light at different wavelengths for $\Delta\omega=0$, i.e.\ the laser frequency lies exactly between the two cavity modes (bottom panel of Fig.\,\ref{fig:cavity}a). We observe that the mode-splitting induces an overall loss in the transmission, which decreases with increasing wavelength. Eventually, the transmission reaches unity at high wavelengths. 

In order to reduce the insertion loss, working at high wavelengths, i.e.\ above 950\,nm, seems obvious. However, as is explained in more detail in Sec.\,\ref{sec:beta}, we are limited by the maximum achievable $\beta$. Given these constraints, we determined 945\,nm to be an ideal wavelength to carry out the experiments. Figure\,\ref{fig:cavity}b shows the cross-(top) and co-polarised (bottom) cavity transmission for this optimised operation-wavelength. The co-polarised data could be reconstructed using the extracted cavity parameters of Fig.\,\ref{fig:cavity}a and Eq.\,\ref{eq:cav_tr}.\newline

\subsection{Characterisation of the quantum dot}
\begin{figure}[h!]
    \centering
    \includegraphics{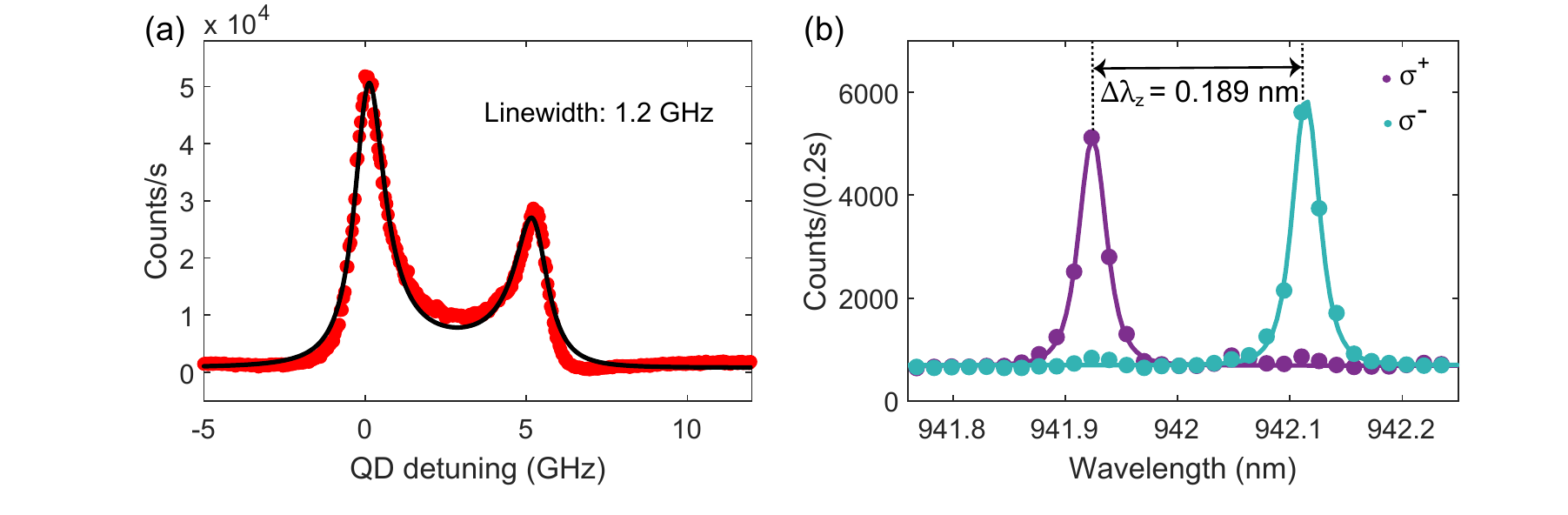}
    \caption{\textbf{Characterisation of the QD} (a) Resonance fluorescence from an over-coupled quantum dot in the microcavity in the cross-polarised configuration. The Purcell-enhanced linewidth is around 1.20\,GHz, corresponding to $\beta = 0.72$. (b)~Photoluminescence spectrum of the quantum dot in a magnetic field of 2.00\,T. A Zeeman splitting of 0.189\,nm is observed. The lower wavelength (purple), i.e.\ higher energy, transition was used for all measurements. The two curves were measured by changing the angle of the quarter-wave plate between the two circular configurations.}
    \label{fig:QD}
\end{figure}

The Purcell factor, $F_p$, is a critical parameter characterising the operation of our diode. It is defined through $\Gamma_{\textrm{tot}} = (1 + F_p)\gamma$, where $\Gamma$ is the overall linewidth of the quantum dot, and $\gamma$ is the linewidth of the quantum dot outside the cavity. The linewidth of a bare quantum dot is typically $\gamma/(2\pi)\approx300$\,MHz, as measured in \cite{tomm_bright_2021}. We estimate the Purcell-enhanced linewidth of the quantum dot by measuring the resonance fluorescence spectrum from the quantum dot. Figure\,\ref{fig:QD}a shows a sample spectrum from a quantum dot placed in the centre of the cavity. We extract $\Gamma_{\textrm{tot}}/(2\pi)=1.2$\,GHz by fitting two Lorentzians to the data. The observed Purcell factor is $F_p = 3$, matching predictions from COMSOL simulations shown in Fig.\,\ref{fig:beta}. 

The two peaks in Fig\,\ref{fig:QD}a show the fine-structure splitting of the optical transition. This splitting between the two transitions increases on applying a magnetic field (\textbf{B}), as seen in Fig.\,\ref{fig:QD}b. At \textbf{B}=2.00\,T, the transitions are split by 63\,GHz (0.189\,nm), large enough that only one of them couples to the laser. The two curves in Fig.\,\ref{fig:QD}b show the photoluminescence (PL) signal of the quantum dot for different positions of the quarter-wave plate (QWP), hence, for different detection polarisations. The purple (turquoise) line represents the QWP setting where the incoming light matches the lower (higher)-wavelength transition. Therefore, only signal from the corresponding transition is collected. The data in Fig.\,\ref{fig:QD} were taken on a different quantum dot to that used in the rest of the paper, but they show the principle of operation.

\newpage
\subsection{Overcoming the dragging effect from the nuclei}
\begin{figure}[h!]
    \centering
    \includegraphics{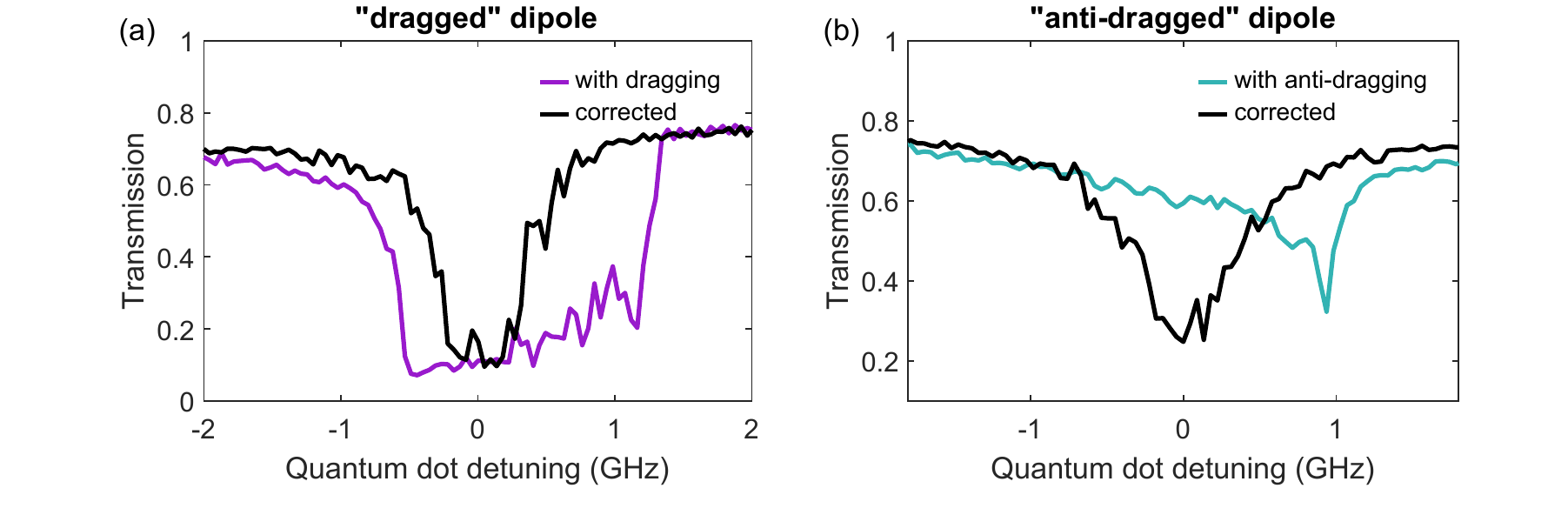}
    \caption{\textbf{Overcoming nuclear spin induced effects} \textbf{(a)/(b)} Transmission through the diode as a function of quantum dot detuning for the high/low energy transition. The so-called ``dragging" and ``anti-dragging" effects (purple/turquoise lines) are overcome (black lines) by jumping back and forth between the measurement voltage and a voltage in the co-tunnelling regime.}
    \label{fig:dragging}
\end{figure}

While operating in a magnetic field, we observed a nuclear-spin related effect in the transmission data, see Fig.\,\ref{fig:dragging}. We found that the transmission depends on the scanning direction of the gate voltage and is asymmetric around the resonance of the quantum dot. In order to overcome these so-called dragging and anti-dragging effects, the gate voltage was switched rapidly (100\,Hz) between the resonance voltage and the co-tunnelling voltage using a square function from an arbitrary waveform generator. In the co-tunnelling regime, an electron is exchanged frequently with another electron, randomising the electron's spin state and eliminating the spin polarisation of the nuclei~\cite{latta_confluence_2009}. Transmission data were acquired only when the gate voltage was at the resonance voltage. The procedure to overcome nuclear spin effects was applied for all transmission data shown in the main text and in the remainder of the supplementary information.

\newpage

\section{Dependence of the $\beta$-factor on the wavelength and the position of the quantum dot}\label{sec:beta}
\subsection{The wavelength dependence}
The coupling efficiency between the quantum dot and the cavity mode is given by the $\beta$-factor which has a wavelength dependence arising from the dispersion of the top and the bottom mirrors. We simulate this dependence by calculating the mode volume and the quality factor of the cavity in COMSOL. The results are shown in Fig.\,\ref{fig:beta}. The $\beta$-factor drops with increasing wavelength and falls below 0.5 at a wavelength of 955\,nm. This effect arises from the fact that the centre of the stop-band for the top and the bottom mirrors are 920\,nm. Away from 920\,nm the reflection of the top mirror drops and results in a reduced quality factor and, hence, a reduction in the $\beta$-factor.

\subsection{The wavelength for the diode operation}
\begin{figure}[h!]
    \centering
    \includegraphics{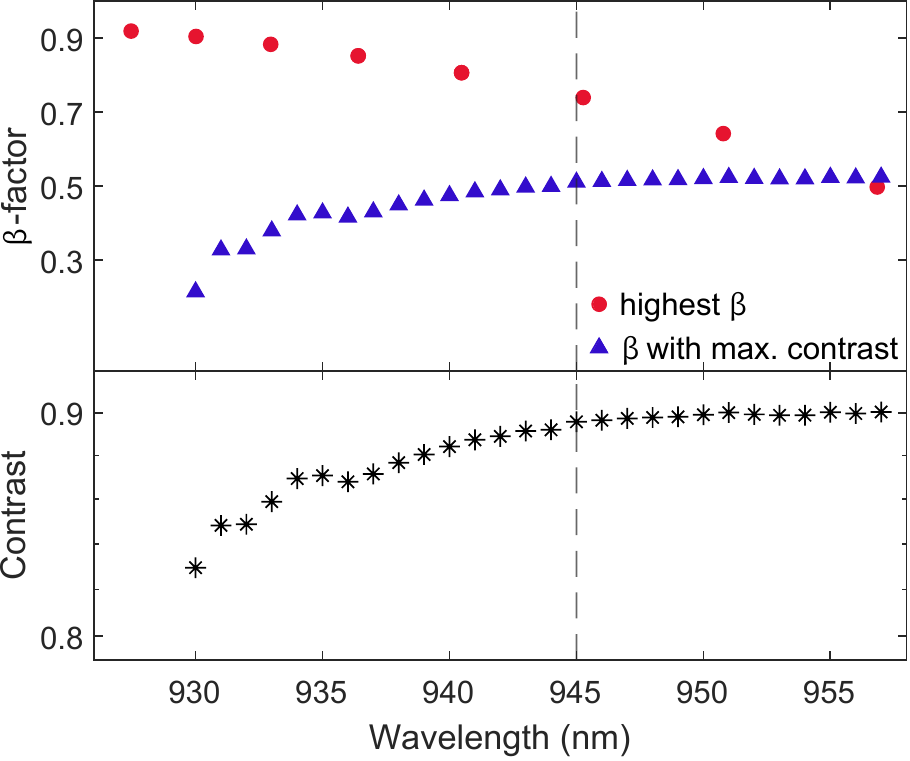}
    \caption{\textbf{Wavelength dependence of the $\beta$-factor.} COMSOL simulations (red circles) show the maximum achievable $\beta$ for a quantum dot positioned in the centre of the cavity. At wavelengths above 955\,nm, the critical $\beta$ of 0.5 is not reachable. Calculations using the model in Sec.\,\RNum{1} enable us to predict the maximal contrast achievable for different wavelengths (black stars) and the optimal $\beta$-factor to observe the strongest contrast (blue triangles). The range between 945 and 955 has the highest contrast at $\beta = 0.5$ and is therefore ideal. The experiments in the main text were carried out at 945\,nm (vertical line).}
    \label{fig:beta}
\end{figure}

Using the model derived in Sec.\,\RNum{1}, $\Lambda$ and $\kappa$ extracted in Sec. \ref{sec:modesplitting}, and leaving $\beta$ a free parameter, we estimate the highest observable contrast ($1-T^0/T^\infty$) for each wavelength (bottom part of Fig.\,\ref{fig:beta}) and extract the $\beta$ at which it is maximised (top panel in Fig.\,\ref{fig:beta}). The achievable contrast is nearly constant between 945 and 955\,nm, giving a broad range of quantum dots to work with. This also shows that the insertion losses discussed in Sec.\,II do not influence the isolation, i.e.\ non-reciprocity, of the system, but only describe overall losses. We chose an emitter at 945\,nm, as the maximum $\beta$ is above the critical value of 0.5, and the position dependence of $\beta$ can be exploited to tune the system into the critical coupling conditions.

\newpage 

\subsection{The position dependence}
The cavity mode has a Gaussian intensity profile centred around the crater. The coupling between the cavity and the quantum dot is maximum when the quantum dot is placed at the centre of the cavity mode. Figure.\,\ref{fig:vcvg}a shows the resonance fluorescence of a quantum dot as a function of its lateral position. The origin of the coordinate system corresponds to the centre of the cavity mode. The fluorescence from the quantum dot drops as it is displaced from the origin, which shows that the coupling efficiency between the cavity and the quantum dot is reduced. This position dependence can be used to achieve $\beta=0.5$ with high precision as shown in the main text. The position with $\beta=0.5$, used in most measurements in the main text, is indicated by the blue circle in Fig.\,\ref{fig:vcvg}a. 

\subsection{Over-coupled cavity-quantum dot system}
\begin{figure}[h!]
    \includegraphics{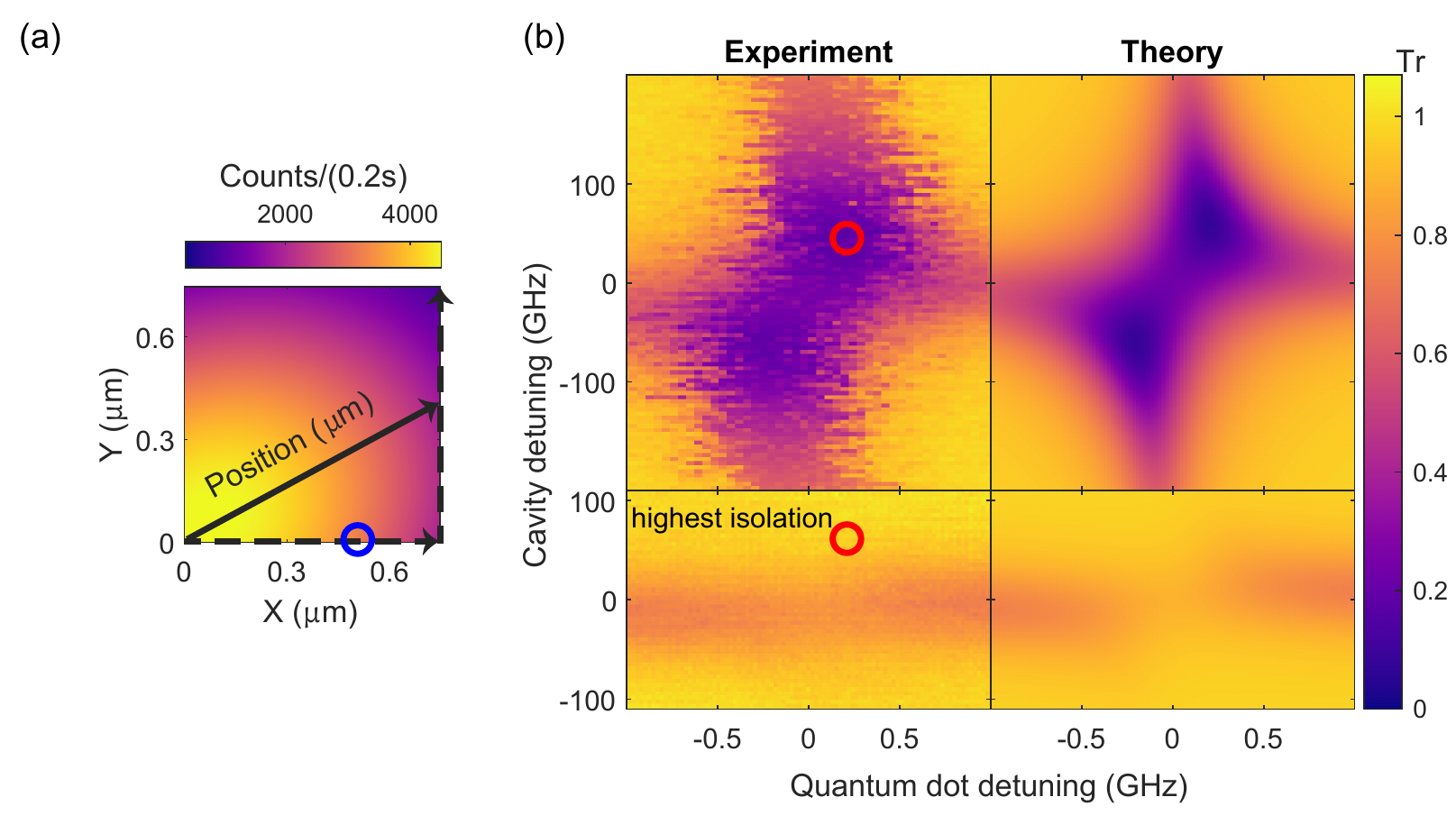}
    \caption{(a) Intensity of resonance fluorescence as a function of the displacement of the quantum dot from the cavity mode centre. The position is defined as the distance from the anti-node with respect to the top mirror of the cavity, i.e.\ $r = \sqrt{(X^2+Y^2)}$. We apply a bias to the X and Y nanopositioners to move the semiconductor sample. The applied voltages are translated to distance using the specifications from the manufacturer (attocube systems AG). The position of the quantum dot for most of the measurements in the main text (highest transmission contrast) is indicated with a blue circle. (b)~Transmission as a function of cavity and quantum dot detuning for the same quantum dot as in the main text, but in the centre of the cavity ($r=0$) where $\beta=0.72$. The strongest contrast is 0.89 which appears at a spectral detuning of 64\,GHz from the cavity resonance (red circles). The model explains the data very well. }
    \label{fig:vcvg}
\end{figure}

Figure\,\ref{fig:vcvg}b shows the transmission as a function of cavity and quantum dot detuning, i.e.\ the same as Fig.\,2a,b in the main text, for an over-coupled quantum dot (the same quantum dot but now positioned in the centre of the cavity). For this situation, the strongest contrast appears when the quantum dot is spectrally detuned from the cavity resonance (contrast 0.89, red circles); the Purcell factor, $F_p$, hence also $\beta$, decrease as a function of cavity detuning. In this case, the isolation resulting from the detuned quantum dot-cavity system is 11.2 (10.5\,dB), which is similar to the value reported in the main text (10.7\,dB). The panels on the right side of Fig.\,\ref{fig:vcvg} are reproduced using the model in Sec.\,I, with $\beta=0.72$, $\kappa/(2\pi)=102$\,GHz, $\Lambda/(2\pi)=29$\,GHz, and a spectral fluctuation of $\delta_{\textrm{sf}}/(2\pi)=35$\,MHz. The theoretical plots in Fig.\,2 of the main text are reproduced with the same parameters but with $\beta=0.5$.

\newpage

\section{Second-order correlation function}
\begin{figure}[h!]
    \centering
    \includegraphics{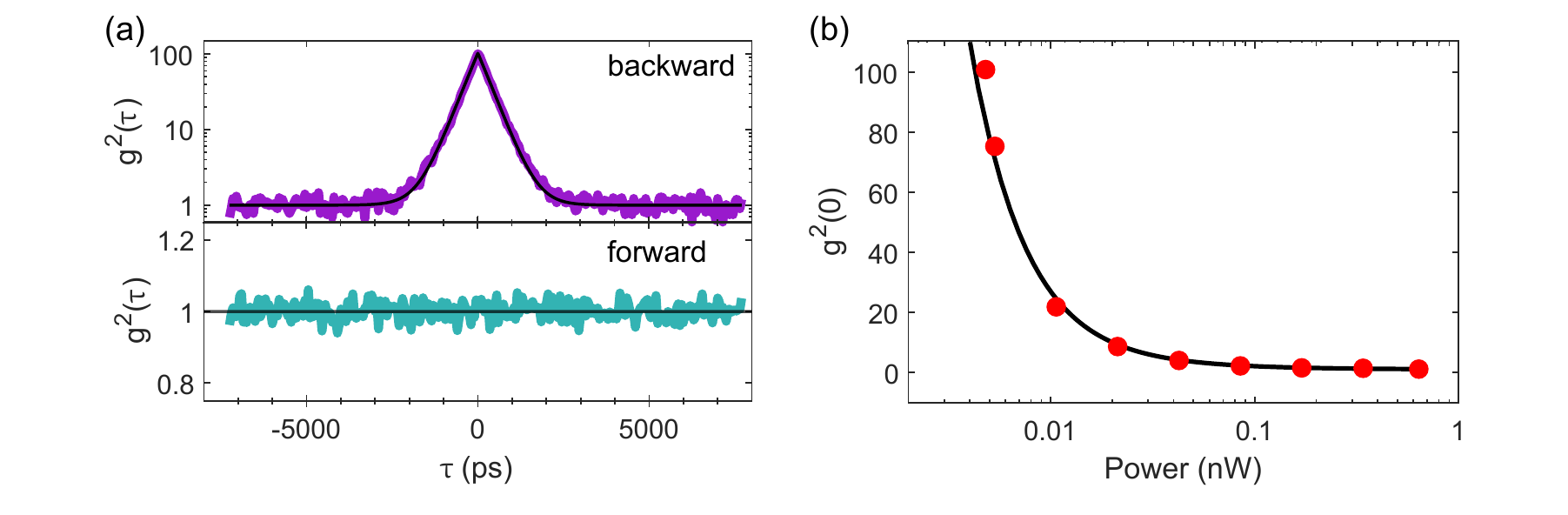}
    \caption{(a) Auto-correlation function $g^{(2)}(\tau)$ for a critically-coupled quantum dot-cavity system for the two propagation directions. Top: Backward direction showing a high bunching of 101. Bottom: Forward direction showing a flat curve indicating that in the forward direction there is no interaction between the photons and the quantum dot such that the light remains coherent after passing through the cavity. The input power is around 5\,pW. The black, solid lines are based on the model in Eq.\,19. (b) $g^{(2)}(0)$ as a function of power in the backward direction. The observed bunching behaviour is a very strong function of power. The solid line is based on the model in Eq.\,\ref{eq:g2pow}.}
    \label{fig:g2}
\end{figure}

We measured the intensity auto-correlation function $g^{(2)}(\tau)$ as a function of time delay $\tau$ for the photons propagating in the forward (port 1 to port 2) and in the backward (port 2 to port 1) directions. The recorded data for both directions at the lowest power is shown in Fig.\,\ref{fig:g2}a. $g^{(2)}(\tau)$ in the forward direction is a flat line, corresponding to that of coherent light, as there is no interaction with the emitter in the forward direction and the light's photon statistics are not affected. In the backward direction, a $g^{(2)}(\tau = 0) = 101$ is observed at zero delay and low power of 5\,pW. This bunching decays exponentially at higher power as can be observed in Fig.\,\ref{fig:g2}b. 

In order to model $g^{(2)}(\tau)$ of the backwards-propagating photons, we base ourselves on the derivation of Ref.~\cite{rice_single-atom_1988}. A comparison of the equations of motion in Ref.~\cite{rice_single-atom_1988} to the ones in Eq.\,\ref{eq:refl_tot} (in resonance condition: $\Delta\omega=0$, $\delta_{\textrm{H/V}}=0$ $\rightarrow$ $t_{H/V}=1$) reveals that the equations of motion for a two-sided (2S) cavity are the same as the equations of motion for a one-sided (1S) cavity under the transformation:
\begin{equation}
    \frac{1}{1/F_{\textrm{p,2S}}+1} = \frac{2}{1/F_{\textrm{p,1S}}+1},
    \label{eq: relation_betas}
\end{equation}
where $F_{\textrm{p,2S}}$ is the Purcell factor as defined in Ref.~\cite{rice_single-atom_1988} and $F_{\textrm{p,1S}}$ is the Purcell factor in Eq.\,\ref{eq:refl_tot} (labelled $F_p$ in Sec.\,I-III). This transformation is equivalent to transforming the $\beta-$factor as $\beta_{\textrm{2S}}=2\beta_{\textrm{1S}}$. Note that this substitution can also be seen from the difference in the simplified transmission equations, which for a two-sided cavity is $T=|1-\beta|^2$ and $T=|1-2\beta|^2$ for a one-sided cavity.

We use the result for $g^{(2)}(\tau)$ in Ref.~\cite{rice_single-atom_1988} to model $g^{(2)}(\tau)$ in our experiments. The full formula for $g^{(2)}(\tau)$ is given by      
\begin{equation}
    g^{(2)}(\tau) = 1 + \left(\frac{\sqrt{2}F_{\textrm{p,2S}}}{1+ Y^2}\right)^2 \cdot~ e^{(3/4)\tau'} \cdot \Bigg\{ \left(Y^2 + \frac{F_{\textrm{p,2S}}^2}{2} -1 \right) \cdot \cosh(\Omega \tau')+\frac{1}{4\Omega}\cdot \left[ Y^2 \cdot \frac{5-F_{\textrm{p,2S}}}{1+F_{\textrm{p,2S}}} -1 - \frac{F_{\textrm{p,2S}}^2}{2} \right] \cdot \sinh(\Omega \tau') \Bigg\},
    \label{eq:full_g2}
\end{equation}
with $\Omega = 1/4 \cdot \sqrt{1- (8Y)^2/((1+F_{\textrm{p,2S}})^2}$; $Y$ is the Rabi angular frequency in units of $\gamma$ ($\textrm{Y}=\sqrt{\Gamma}/\sqrt{\gamma}\cdot b_{\textrm{in}}$); and $\tau'$ is a dimensionless delay, $\tau'\,=\,\gamma(1~+~F_{\textrm{p,2S}})\cdot~\tau$.
In the low power limit ($Y\ll 1$), this reduces to:
\begin{equation}
    g^{(2)}(\tau) = (1-F_{\textrm{p,2S}} ^2 \cdot e^{-\tau'/2})^2, 
    \label{eq:g2}
\end{equation}
and
\begin{equation}
    g^{(2)}(0) = (1-F^2_{\textrm{p,2S}}) ^2 . 
    \label{eq:g20}
\end{equation}
The power dependence of the bunching is modelled by setting the time delay $\tau = 0$, resulting in:
\begin{equation}    
g^{(2)}(0) = 1 + \left(\frac{\sqrt{2}F_{p,2S}}{1+ Y^2}\right)^2 \cdot \left(Y^2 + \frac{F_{p,2S}^2}{2} -1 \right).
\label{eq:g2pow}
\end{equation}
These equations describe our measurements very well. Note that all the modelling for the correlation function was carried out for the case of a degenerate cavity for simplicity. We use $F_{\textrm{p,1S}} = 0.8$ to model our results in the backward direction. The discrepancy with the expected value of $F_{\textrm{p,1S}} = 1$ likely results from the residual spectral fluctuation (see last part of Sec.\,I and Eq.\,\ref{eq:diff}) as $g^{(2)}(\tau)$ has a strong dependence on dephasing at low powers \cite{lodahl_interfacing_2015}. 

\newpage

\newpage

\bibliography{nonreciprocity}

\end{document}